\documentclass{article}

\usepackage{arxiv}

\usepackage[utf8]{inputenc} 
\usepackage[T1]{fontenc}    
\usepackage{hyperref}       
\usepackage{url}            
\usepackage{booktabs}       
\usepackage{amsfonts}       
\usepackage{nicefrac}       
\usepackage{microtype}      
\usepackage{cleveref}       
\usepackage{lipsum}         
\usepackage{graphicx}
\usepackage{natbib}
\usepackage{doi}
\usepackage{setspace}

\usepackage{graphicx}  
\usepackage{subcaption} 

\usepackage{tikz}

\title{A Method for Rapid 
Area Prioritisation 
in\\Flood Disaster Response}

\newif\ifuniqueAffiliation
\uniqueAffiliationtrue

\usepackage{authblk}

\newbox{\orcid}\sbox{\orcid}{\includegraphics[scale=0.06]{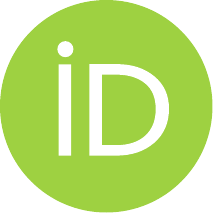}} 
\author[1]{%
    {\hspace{1mm}Moritz~Schneider\thanks{\texttt{Corresponding Author, e-mail:~moritz.schneider@dlr.de}}}%
}
\author[1]{%
	{\hspace{1mm}Lukas~Halekotte}%
}
\author[2]{%
	{\hspace{1mm}Tina~Comes}%
}
\author[3]{%
	{\hspace{1mm}Frank~Fiedrich}%
}
\affil[1]{Institute for the Protection of Terrestrial Infrastructures, German Aerospace Center, Sankt Augustin, Germany}
\affil[2]{Faculty of Technology, Policy and Management, TU Delft, Delft, The Netherlands}
\affil[3]{Chair for Public Safety and Emergency Management, University of Wuppertal, Wuppertal, Germany}

\renewcommand{\headeright}{Prioritisation Recommendation Mapping (PrioReMap)}
\renewcommand{\shorttitle}{} 

\hypersetup{
pdftitle={A template for the arxiv style},
pdfsubject={q-bio.NC, q-bio.QM},
pdfauthor={David S.~Hippocampus, Elias D.~Striatum}, 
pdfkeywords={First keyword, Second keyword, More},
}

\begin{document}
\maketitle

\begin{abstract}
In flood disasters, decision-makers have to rapidly prioritise the areas that need assistance based on a high volume of information.
While approaches that combine GIS with Bayesian networks are generally effective in integrating multiple spatial variables and can thus reduce cognitive load, existing models in the literature are not equipped to address the time pressure and information-scape that is typical in a flood.
To address the lack of a model for area prioritisation in flood disaster response, we present a novel decision support system that adheres to the time and information characteristics of an ongoing flood to infer the areas with the highest risk. 
This decision support system is based on a novel GIS-informed Bayesian network model that reflects the challenges of decision-making for area prioritisation.
By developing the model during the preparedness phase, some of the most time-consuming aspects of the decision-making process are removed from the time-critical response phase.
In this way, the proposed method aims to 
providing rapid and transparent area prioritisation recommendations for disaster response.
To illustrate our method, we present a case study of an extreme flood scenario in Cologne, Germany. 
\end{abstract}

\keywords{Spatial Prioritisation, Flood Management, Disaster Response, Decision Support, Bayesian Network, GIS.}

\section{Introduction}
Flooding is among the most common and devastating natural hazards, posing severe threats to human life and critical infrastructures \citep{Nick2023,Asaridis2025}. Due to climate change and urbanisation, the flood risk for urban areas has substantially grown in the recent past and is very likely to grow further in the future \citep{Qi2022}.
Therefore, cities around the globe need to prepare for increasing flood risk. 
A fast and targeted response to a flood requires a rapid yet comprehensive assessment of the situation.
One crucial aspect of this evaluation is the identification and prioritisation of the most impacted areas in the disaster zone, which helps guide rapid response efforts and efficiently allocate limited resources \citep{Armenakis2012, Wieland2025}.

In the initial stages of a flood disaster, the actual impact of the event is often not exactly known or not yet fully developed but needs to be estimated based on proxy data that is available.
Effectively, this means that operational impact assessments are risk-based -- they estimate where the impact is expected to be highest.
Since disaster risk results from the intersection of hazard, exposure, and vulnerability \citep{Lu2024, Mentges2023}, it is necessary to consider these three dimensions when identifying and prioritising the at-risk areas. This includes, for instance, the spatial distribution of the hazard intensity \citep{Mudashiru2021}, the density of residential buildings, the accessibility of critical infrastructures (CI) \citep{casali2024data}, and the location of vulnerable people \citep{Rehman2019}. Especially at the onset of a flood, these variables tend to be uncertain \citep{Schneider2025}, particularly in terms of human adaptive behaviour \cite{sirenko2024rhythm}. 

Despite uncertainties and many information gaps, decision-makers are at the same time confronted with a high volume and velocity of information. This combination has been shown to result in cognitive overload on decision-makers that can hamper their capacity to manage disaster response efforts efficiently \citep{van2016improving, Comes2024}.
To reduce the cognitive load on decision-makers and to guide them in prioritising at-risk areas, suitable decision support systems are required that consider the available information, rapidly process it, and present it in a concise manner.

In the literature, a combination of geoinformation systems (GIS) and Bayesian networks (BNs) has been proven to be a powerful tool to support \textit{long-term} decisions in flood disaster risk reduction, e.g. for informing flood mitigation by identifying flood prone areas \citep{Zwirglmaier2024,Wu2019}, or flood protection by assessing potential CI failures \citep{Schneider2025a,KantiSen2020}.  
By combining a GIS with a BN, area-specific assessments that include a variety of spatially distributed variables can be conducted (e.g. see \cite{Lu2024} or \cite{Zwirglmaier2024}).
In this context, the popularity of BNs stems from three aspects that are particularly critical in disaster-related contexts: (i) BNs can be constructed based on diverse compositions of available data, e.g. a combination of historical data and expert knowledge \citep{Druzdzel2000}, (ii) BNs are easy to interpret as they are based on a (logical) graphical structure \citep{lee2022roadmap}, and (iii) BNs enable the consideration of uncertainties in input data and relations between variables \citep{pearl1985bayesian} -- which is crucial in high-stake decisions to avoid overconfidence \citep{Schneider2025}.

The existing GIS-informed BN models are designed for informing strategic decisions in disaster risk reduction. 
As a result, they are not equipped to support decisions in flood disaster response, which are characterized by the tension between the need for rapid decisions and the complexity of the situation. 
However, BN-based approaches are generally well suited to address the particularities of disaster situations. 
A BN-based approach, designed to reflect the cognitive process of disaster response decisions, can generally be utilized to treat some of the more time-consuming aspects of the decision-making process during the model development and thus draw them out of the time-critical disaster response phase.
This involves addressing questions regarding (i) the variables that are critical for disaster risk, (ii) the dependencies between these variables, (iii) the level of risk that justifies area prioritisation, and (iv) the treatment of uncertainties in the available information. 
Properly treating these questions is a non-trivial and time-consuming process that may require different needs to be taken into account and thus involve consulting multiple experts \citep{Zheng2020}. 
Accordingly, holding appropriate discussions when time permits and moulding the results into a comprehensible model (that is the BN) can help decision makers to make faster, better justified, and thus more confident decisions when time presses.

In this work, we present a novel method called \textit{PrioReMap} (Prioritisation Recommendation Mapping) to support flood response decisions based on a GIS-informed BN model.
This method includes two layers that together constitute the contribution of our work: 
(i) a novel GIS-informed BN model that is tailored to the information-scape in an ongoing flood disaster; and 
(ii) a novel area prioritisation method, with the GIS-informed BN model at its core.
The resulting recommendations can help decision-makers allocate scarce resources to priority areas or to select appropriate locations for temporary storage, shelters, field hospitals, or command centres.

In the remainder of this paper, we first provide the background on GIS-informed BN models, especially in the context of flood disasters (see Section \ref{studies}). 
Based on the literature, we derive design properties of such models. 
We demonstrate the need for a new approach based on these design properties via a stylised case (see Section \ref{motivation}).
Subsequently, we present our method for area prioritisation (see Section \ref{method}). 
We illustrate the effectiveness of our \textit{PrioReMap} method in a case study of a flood scenario in Cologne, Germany (see Section \ref{case_study}).
Finally, the proposed method is discussed and future work is outlined (see Section \ref{discussion}).

\section{Background: GIS-informed Bayesian Networks} \label{studies}
The combination of GIS and Bayesian networks has become increasingly popular for spatial inference. 
For instance, GIS-informed BN have been applied to optimise the placement of charging stations for electric vehicles \citep{Zhang2022}, to find the most suitable sites for pumped hydro energy storage \citep{Ali2024}), to evaluate the suitability of underground space as a resource for urban development \citep{Xu2023}, or to identify suitable areas for timber production under conservation constraints \citep{GonzalezRedin2016}. In disaster risk reduction, this approach has frequently been applied for creating risk maps, e.g. for fire  \citep{Dlamini2010}, avalanches \citep{GretRegamey2006}, or ecological risk \citep{Guo2020}. 
GIS-informed BN models are particularly well established in the context of flood disaster risk reduction (see Table \ref{literature} for an overview).

Although the objective of GIS-informed BN can differ, their construction tends to follow three general design properties (DP1-DP3):
\paragraph{DP1 -- Discrete target node represents the analysis objective}
The first step in constructing a GIS-informed BN is the specification of a target node that reflects the objective of the analysis (see Table \ref{literature}). 
For example, \cite{Lu2024} present a BN-based approach to identify areas of high flood disaster risk, which is reflected in the target node \textit{Flood Disaster Risk}.
The possible outcomes of the analysis are then set by the states that this target node can adopt. 
In all applications of GIS-informed BN for flood risk, the target nodes show discrete states, which can be (i) binary states, such as \textit{Yes} and \textit{No} or \textit{True} and \textit{False}, or (ii) higher granularity state descriptions, such as \textit{High}, \textit{Medium}, and \textit{Low} (see Table \ref{literature}). 
The motivation for using discrete instead of continuous states is that models in this context are often primarily built from expert knowledge, and discrete states are often easier to elicit from experts. 

\paragraph{DP2 -- Study area is divided into sub-areas}
To develop the GIS models, the study area must be divided into smaller subset areas that are individually assessed.
To achieve this, the area can be divided by using regular tiles, such as squares (e.g. see \citep{Lu2024}) or spatially distributed components, such as buildings (e.g. see \citep{Schneider2025}).
These subset areas determine the resolution of the subsequent analysis.

\paragraph{DP3 -- Leaf nodes are informed by GIS models}
The target node, which is decomposed through one or more layers of parent nodes, leads to independent leaf nodes that are ultimately informed by the area-specific GIS models.
For example, the target node \textit{Flood Disaster Risk} in \cite{Lu2024} has three parent nodes called \textit{Hazard}, \textit{Exposure}, and \textit{Vulnerability}, which, in total, depend on six parent (leaf)nodes. 
Using the example of node \textit{Exposure}, which has two parent (leaf) nodes called \textit{River Network Density} and \textit{River Buffering} that are ultimately informed by two individual GIS models. 
For each individually assessed area, spatially explicit information on all BN leaf nodes must be provided. 
When informing the area-specific BN leaf nodes with geospatial data, two cases of evidence must be distinguished: hard evidence and soft evidence.
Hard (or regular) evidence describes a deterministic value that gives the exact state of a leaf node.
Soft evidence, on the other hand, describes a probability ratio of a leaf node.

\begin{table}[!h]
    \centering
    \caption{GIS-informed BN models for flood disaster preparedness in the literature.}
    \label{literature}
    \resizebox{\textwidth}{!}{\begin{tabular}{|p{0.2\textwidth}|p{0.2\textwidth}|p{0.2\textwidth}|p{0.2\textwidth}|p{0.2\textwidth}|}
    \toprule
        \textbf{References} & \textbf{Analysis Objective} & \textbf{Target Node} & \textbf{Target Node States} & \textbf{BN Leaf Nodes} \\\midrule
        \cite{Schneider2025a} & Analysis of hospital service disruptions & Emergency Care & True, False & Flood Depth at Hospital, Accessibility, Power Supply Grid\\
        \cite{Lu2024} & Flood disaster risk assessment & Flood Disaster Risk & Extreme High, High, Moderate, Low, Very Low & Inundation Extent, River Network Density, River Buffering, Population Vulnerability, Economic Vulnerability, Building Vulnerability \\
        \cite{Zwirglmaier2024} & Integrated flood risk assessment in data-scarce mega-cities & Flood Hazard Probability & Yes, No & Remote Sensed Urban Structure Types, Proximity to Large Scale Green Infrastructure, Topography, Proximity to River, Proximity to Coast \\
        \cite{Arango2022} & Flood risk assessment for road infrastructures & Flood Risk Factor & Low, Medium, High & Extreme Precipitation Susceptibility Index, Historical Records, Repair Cost, Light Traffic, Population Density, Soil, River Density \\
        \cite{Huang2021a} & Disaster-causing factor chains on urban flood risk & Inundation & Yes, No & Elevation, Population Density, Annual Rainfall \\
        \cite{KantiSen2020} & Quantify resilience of roadways network infrastructure & Resilience & Low, Medium, High & Reliability, Recovery (of road network components)\\
        \cite{Wu2019} & Assessing urban flood disaster risk & Flood Disaster & Yes, No & River Density, Proximity, Elevation, Impervious Area, Per Unit GDP, Road Density, Population Density, Rainfall Duration \\
        \cite{Balbi2016} & Assessing benefits of early warning systems & Vulnerability & Low, Medium, High & Emergency Personnel, People Risk Awareness, Reliability \\
\bottomrule
    \end{tabular}}
\end{table}

\section{Motivation - From Layers to Decision Support} \label{motivation} 
Based on the aforementioned basic design properties (\textbf{DP1-DP3}) of GIS-informed BN models in the literature (see Section \ref{studies}), we construct a simple (low-dimensional) example model to illustrate how even a basic example yields results that can become difficult to comprehend -- a critical issue especially under time pressure. 

\begin{figure}[ht]
    \centering
    \includegraphics[width=1.1\textwidth]{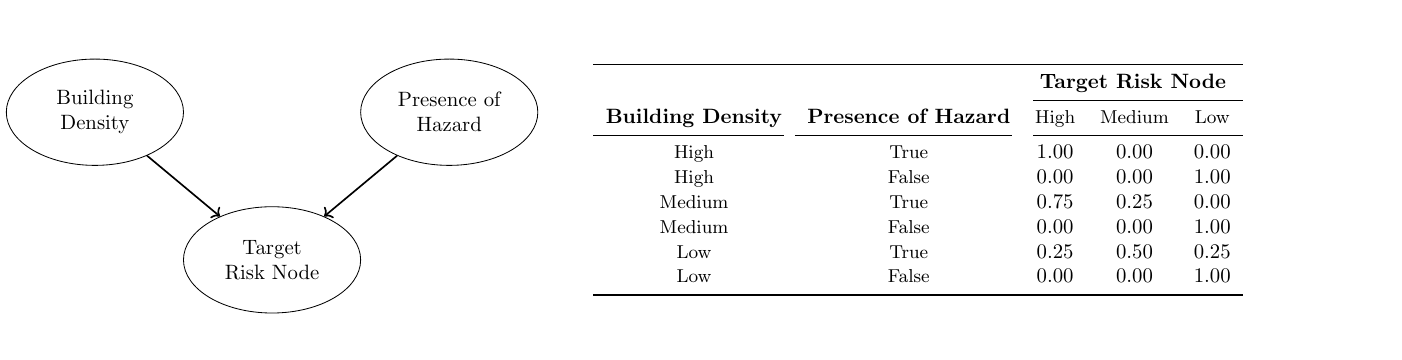}
    \caption{Low-dimensional BN model composed of the directed acyclic graph (left side) and CPT (right side).}
    \label{fig:bn_example}
\end{figure}

\begin{figure}[ht]
    \centering
    \begin{subfigure}[b]{0.25\textwidth}
        \includegraphics[width=\textwidth]{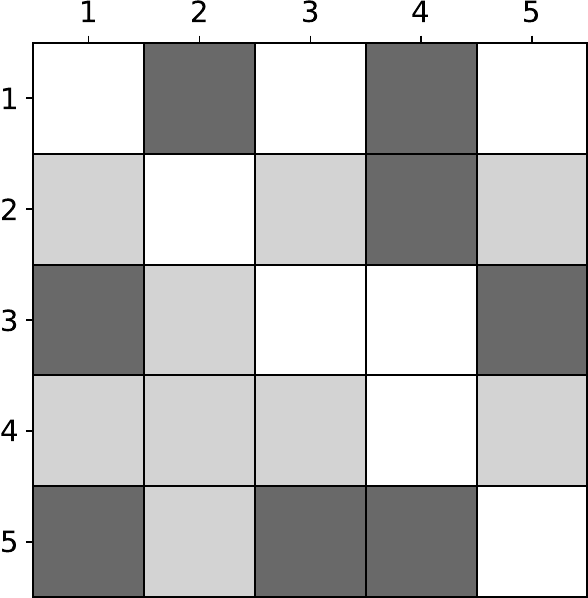}
        \caption{Homogeneous spatial data.}
        \label{fig:node1}
    \end{subfigure}
    \hspace{1cm}
    \begin{subfigure}[b]{0.25\textwidth}
        \includegraphics[width=\textwidth]{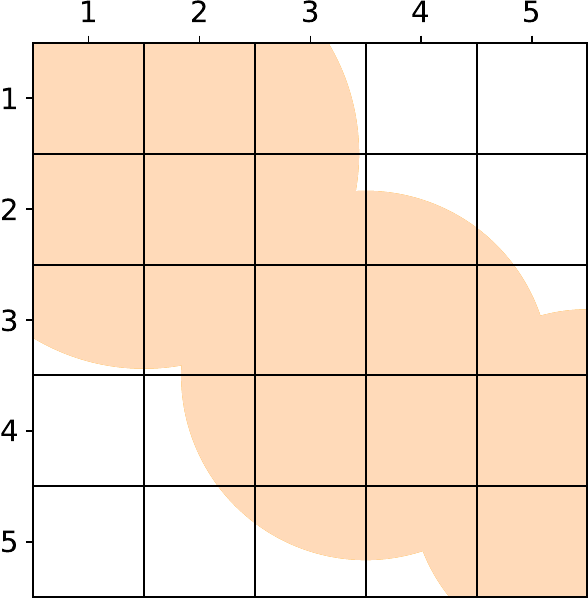}
        \caption{Heterogeneous spatial data.}
        \label{fig:node2}
    \end{subfigure}
    \caption{Example GIS-data on a $5x5$ matrix. Fig. \ref{fig:node1} shows (homogeneous) building density geo-data in each cells with three configurations (high density in dark-gray, medium density in light-gray, and low density in white) processed as hard evidence. Fig. \ref{fig:node2} shows geo-data that is not bound to tight to the boundary of the cell processed as soft evidence.}
    \label{fig:two_images}
\end{figure}

The BN of this example features a \textit{Target Risk Node} with three states (\textit{High, Medium}, and \textit{Low}) (following \textbf{DP1}) and two parent nodes (see left side of Figure \ref{fig:bn_example}): (i) a node \textit{Building Density} with three states (\textit{High}, \textit{Medium}, and \textit{Low}); 
and (ii) a node \textit{Presence of Hazard} with two states (\textit{True} and \textit{False}). 
The corresponding conditional probability table (CPT) attached to the \textit{Target Risk Node} follows a simple structure: Given the presence of the hazard, cells with a high building density are most likely to obtain a high risk, followed by those with a medium building density, and those with a low building density (see right side of Figure \ref{fig:bn_example}). 
As a geographical setup, we introduce a $5\times5$ matrix (see Figure \ref{fig:two_images}) with $25$ cells (following \textbf{DP2}). Each cell of the matrix represents one subset area, i.e. the BN is duplicated and assigned to each cell. 
The two leaf nodes of each BN are informed by individual GIS layers (following \textbf{DP3}).
The node \textit{Building Density} is informed by a layer containing geo-data of the density of residential buildings that is assumed to be homogeneous in each cell (see Figure \ref{fig:node1}) and processed as hard evidence in the BN. 
The node \textit{Presence of Hazard} is informed by a layer containing geo-data of the hazard distribution that is not tight to the boundaries of the cells (not homogeneous in every cell, see Figure \ref{fig:node2}) and processed as soft evidence in the BN.

\begin{figure}[ht]
    \centering
    \includegraphics[width=.9\textwidth]{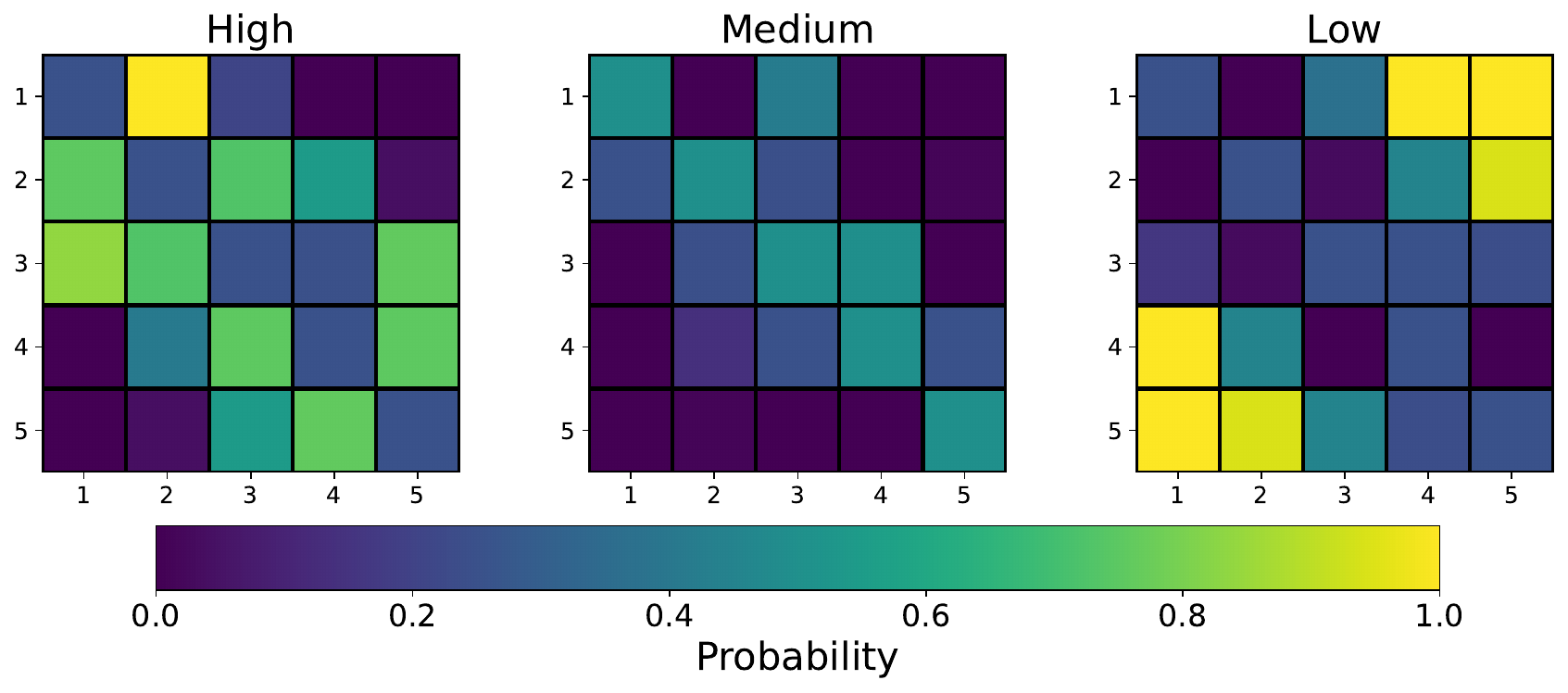}
    \caption{Heatmaps presenting the state probabilities of the \textit{Target Risk Node} (see Figure \ref{fig:bn_example}) calculated using the cell-specific geo-data (see Figure \ref{fig:two_images}) to inform the cell-specific BN leaf nodes.}
    \label{fig:plots_example}
\end{figure}

The GIS-informed BN model provides probabilities for each cell to be in a state of \textit{High}, \textit{Medium}, or \textit{Low} risk.
Based on the corresponding heatmaps (Fig. \ref{fig:plots_example}), tendencies of high risk areas can easily be identified, especially in comparison to having only the two layers of geo-data (see Fig. \ref{fig:two_images}). 
Accordingly, the model is already capable of reducing the cognitive load on potential decision makers. 

Looking more closely at the results, they reveal one cell that stands out with a $100\%$ probability of high risk (see cell in row one, column two, i.e. cell (1,2), in Fig. \ref{fig:plots_example}).
While this cell should clearly be prioritised when considering only the high risk matrix, cells (3,5) and (4,5) also show similarly high probabilities of high risk ($\sim80\%$). However, the two cells differ in their remaining probability distribution: in cell (3,5), the remaining $\sim20\%$ is assigned to low risk, whereas in cell (4,5), it is assigned to medium risk. 
As a result, the probability distribution of cell (4,5) can be considered more critical, despite having the same probability for the most severe (\textit{High}) state (\textit{Issue I}) -- this example underscores the importance of taking the entire probability distribution into account when prioritising cells.
A different prioritisation issue becomes apparent when comparing cells (2,1) and (3,1). 
While cell (3,1) exhibits a slightly higher probability of high risk ($\sim84\%$ vs. $75\%$), cell (2,1) shows a higher probability of medium risk ($25\%$ vs. $0\%$). 
As a result, the prioritisation depends on whether emphasis is placed solely on high risk or on a combination of high and medium risk levels, i.e. the prioritisation depends on the decision-maker’s preferences (\textit{Issue II}).

To conclude, while the presented low-dimensional example demonstrates the general suitability of a GIS-informed BN model for area prioritisation, it also highlights issues associated with applying this method in time-critical response efforts.
Even though the entire setup is very simple, quickly comprehending the results (\textit{Issue I}) and choosing a set of cells to prioritise (\textit{Issue II}) is still not straightforward and can lead to a high cognitive load (especially under time pressure) as well as discrepancies in prioritisation among different responders -- this is where our proposed method aims to enhance the literature by offering precise recommendations for area prioritisation.
We argue that these recommendations should be based on the entire probability distribution of the target node (rather than solely on the most severe state), in order to further reduce cognitive load (thus avoiding \textit{Issue I}) and to promote consistency in recommendation outcomes across responders by introducing a transparent weighting scheme for the individual target node states (thus avoiding \textit{Issue II}).

\section{PrioReMap Method: Prioritisation Recommendation Mapping} \label{method}
\begin{figure}[ht]
     \centering
     \includegraphics[width=0.85\textwidth]{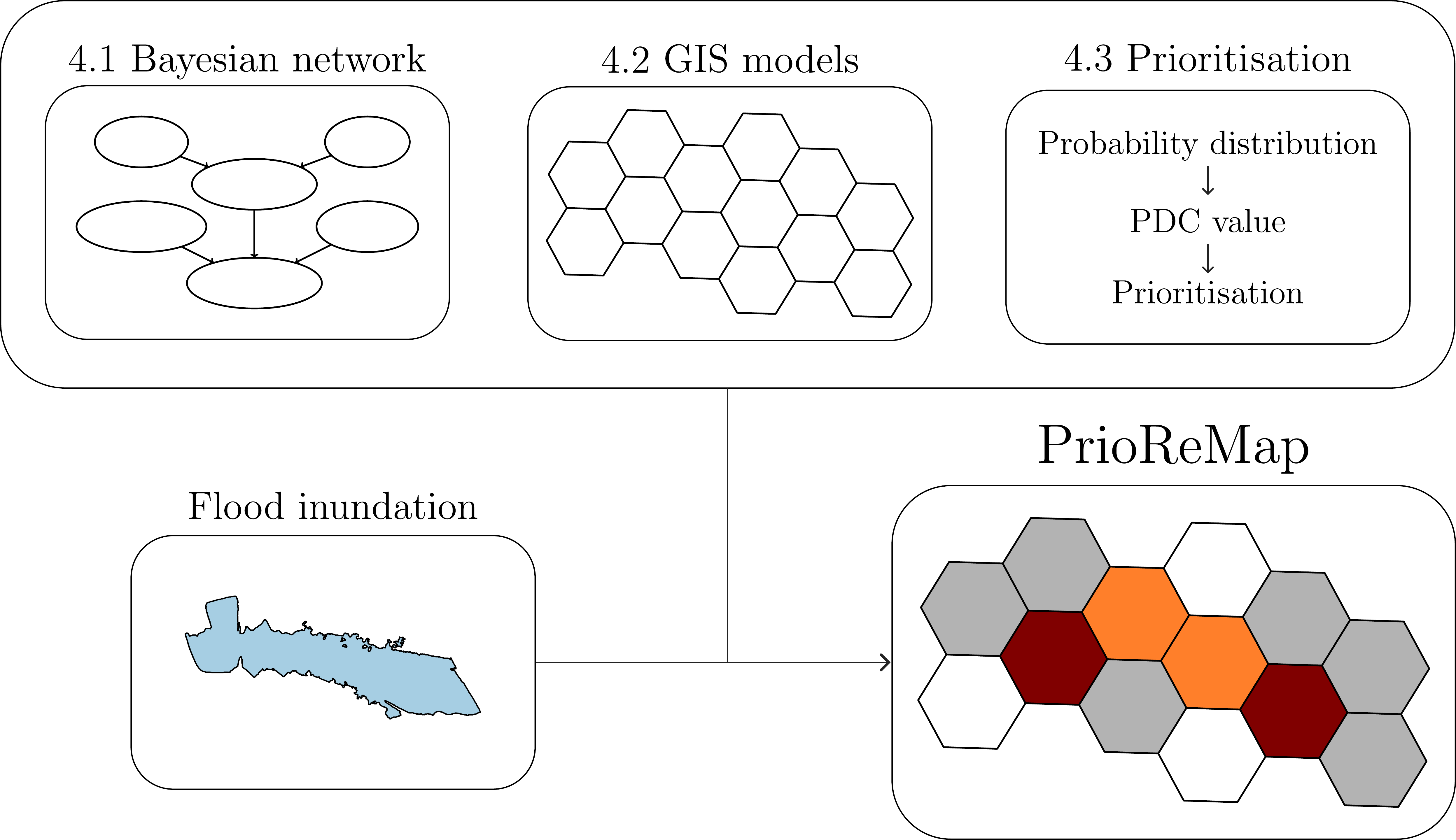}
    \caption{Summary of the \textit{PrioReMap} method.}
     \label{fig:summary_method}
\end{figure}

The method presented here is designed to support responders in prioritising areas in an ongoing flood disaster where the risk to people is high. 
It is composed of three main components (see upper half of Figure \ref{fig:summary_method}): 
(i) the Bayesian network used to infer the probability of the target node called \textit{Risk of People in Need of Assistance} (see Section \ref{bn_model}); 
(ii) the GIS models used to inform the BN leaf nodes with area-specific spatial data (see Section \ref{gis_models}); and 
(iii) the prioritisation method that translates the probability distribution of the target node into prioritisation recommendations (see Section \ref{recommend_method}).
All models and parameters of the method can be set up before a flood event, i.e. during a period without time pressure, which enables a quick recommendation retrieval during the time-critical response phase of a flood event.
However, if prioritisation preferences change during the response phase, the corresponding parameters can still be adjusted.
When applying the method in disaster response, information about the current flood is fed into the corresponding GIS and then processed by the BN and the state weighting procedure to compute the area prioritisation recommendations (see lower half of Figure \ref{fig:summary_method}).

\subsection{Bayesian network} \label{bn_model}
\begin{figure}[ht]
    \centering
    \includegraphics[width=0.6\textwidth]{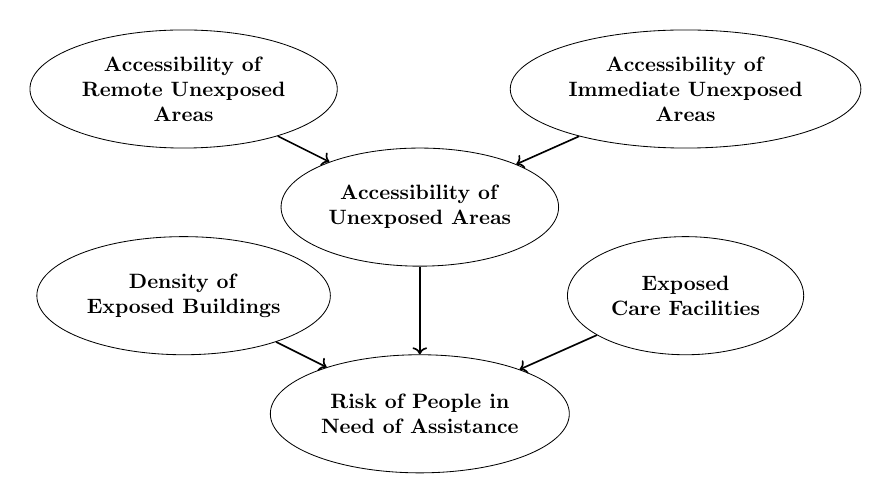}
    \caption{Bayesian network to infer the \textit{Risk of People in Need of Assistance}. The BN is composed of six variables (two conditional nodes and four marginal nodes) that show two to four states (see Table \ref{states}).}
    \label{fig:bn}
\end{figure}

The target node of the BN (see Figure \ref{fig:bn}) depicts the \textit{Risk of People in Need of Assistance} in an individual (subset) area, which corresponds to the most crucial piece of information guiding disaster response efforts \citep{Matsuki2023,SkoeldGustafsson2023}. We selected a set of four possible states (\textit{High}, \textit{Medium}, \textit{Low}, \textit{None}; see Table \ref{states} for all possible node states) to strike a balance between sufficient descriptive granularity and a reasonable effort for filling in the corresponding CPT (an issue that is of particular significance when the corresponding knowledge is elicited from domain experts \citep{Balaram2008}). It is the probability distribution of the target node that is ultimately used to prioritise areas for flood disaster response. The BN is used to assess the target node value by drawing inference from related variables (aka parent nodes). 

We assume that the risk of people in need of assistance depends on the number of people who are exposed to the hazard, which we represent via the proxy variable \textit{Density of Exposed Buildings} in an area \citep{Ehrlich2021} (first parent node). We further assume that the risk is higher if vulnerable individuals reside in the exposed area, which we approximate by the \textit{Presence of Exposed Care Facilities} \citep{Abebe2025} (second parent node). Finally, we assume that the risk in an exposed area decreases if people can escape to an unexposed (safe) area \citep{Li2025}, represented by the node \textit{Accessibility of Unexposed areas} (third parent node). To account for the possibility that people can evacuate towards unexposed areas in their immediate vicinity versus more remote locations, we introduce two additional (second-order) parent nodes -- \textit{Accessibility of Immediate Unexposed Areas} and \textit{Accessibility of Remote Unexposed Areas}.

To both conditional nodes (\textit{Accessibility of Unexposed Areas} and \textit{Risk of People in Need of Assistance}) a CPT is attached. 
The CPT attached to node \textit{Accessibility of Unexposed Areas} follows deterministic rules, i.e. if only one parent node is set to \textit{True}, the \textit{Accessibility of Unexposed Areas} is \textit{Limited}. 
If both parent nodes are in state \textit{True}, the child node is in \textit{True}. Conversely, if both parent nodes are in state \textit{False}, the child node is also in state \textit{False}.
The CPT of the target node is more complex, requiring $96$ probability values, one for each combination of conditional node states and parent node states, i.e. $4x3x2x4=96$.
This CPT is based on the following principles: (i) as the state severity of the node \textit{Density of Exposed Buildings} increases, the probability distribution shifts toward higher probabilities of states \textit{Medium} and \textit{High}; (ii) as the severity of the node \textit{Accessibility of Unexposed Areas} increases, the probability distribution shifts even further toward higher probability of state \textit{High} risk; and (iii) when the node \textit{Presence of Exposed Care Facilities} is in the \textit{True} state, the target node is assigned the \textit{High} state.

\begin{table}[!ht]
    \centering
    \caption{Variables (nodes) and respective states of the BN (see Figure \ref{fig:bn}).}\label{states}
    \resizebox{\textwidth}{!}{\begin{tabular}{|c|c|c|c|c|c|} 
    \hline
    \textbf{Accessibility of} & \textbf{Accessibility of} & \textbf{Accessibility of} & \textbf{Density of} & \textbf{Exposed} & \textbf{Risk of People} \\
    \textbf{Remote Unexposed Areas} & \textbf{Immediate Unexposed Areas} & \textbf{Unexposed Areas} & \textbf{Exposed Buildings} & \textbf{Care Facilities} & \textbf{Need of Assistance} \\
    \hline
    True & True & False   & None  & Present     &  None    \\ \hline
    False & False & Limited & Low    & Not present &  Low     \\ \hline
          &      & True    & Medium &             &  Medium  \\ \hline
          &      &      & High   &             &  High    \\ \hline
    \end{tabular}}
\end{table}

\subsection{GIS models} \label{gis_models}
\begin{figure}[ht]
    \centering
    \includegraphics[width=0.6\textwidth]{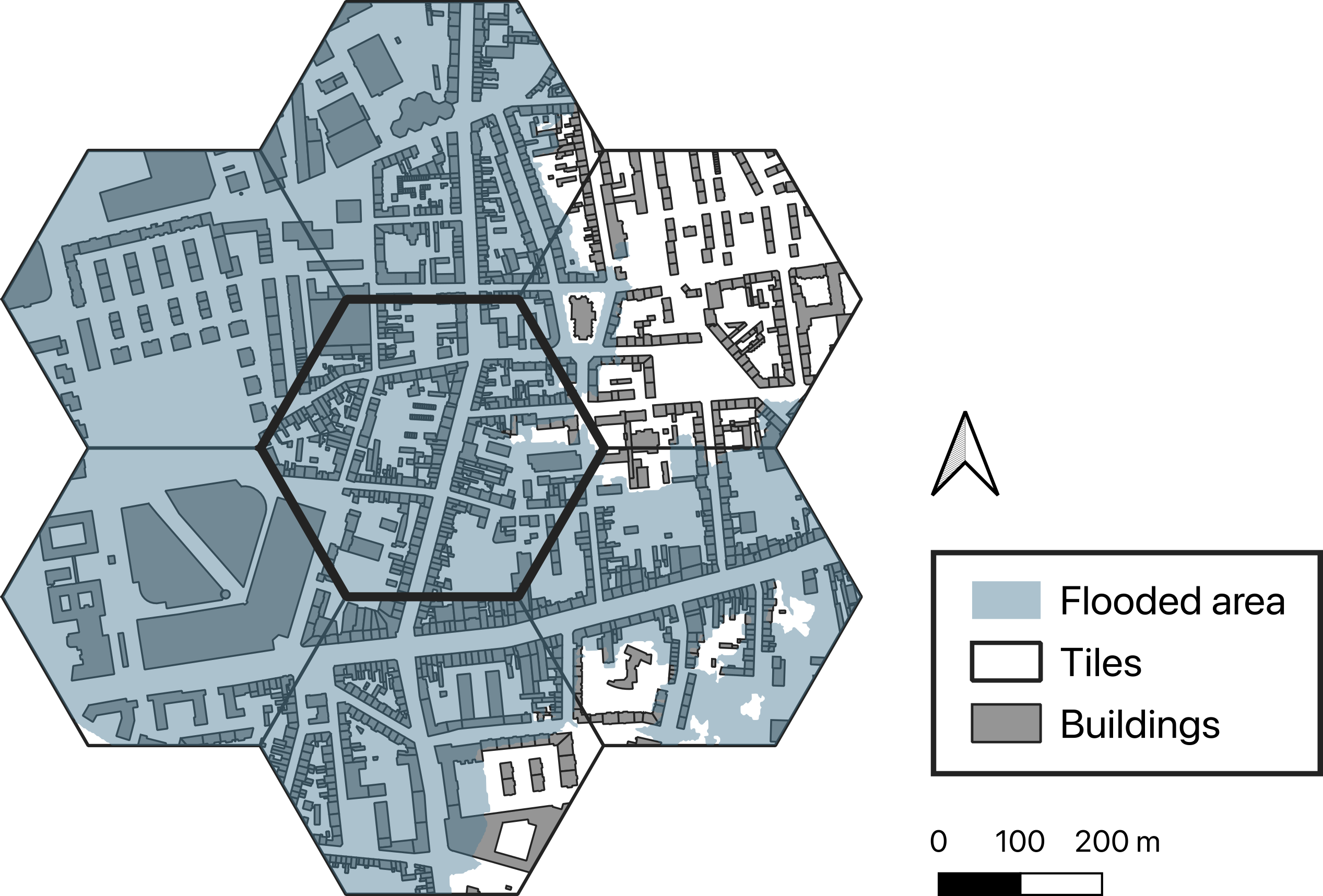}
    \caption{Example hexagon including neighbouring hexagons.}
    \label{fig:single_hex}
\end{figure}

To enable the prioritisation of individual subset areas, i.e. specific zones within the broader disaster area, these areas must first be defined.
To achieve this, the study area is divided into a grid composed of uniform hexagon tiles (see Figure \ref{fig:single_hex}). 
This hexagonal tilling is a common practice in geospatial applications as it ensures uniform neighbour distances, reduced edge effects, and a consistent spatial representation (see \cite{sahr2011hexagonal}).
To create area-specific recommendations, a duplicate of the BN (Figure \ref{fig:bn}) is assigned to each tile and each of its four BN leaf nodes is informed by tile-specific information from an individual GIS model. 
Each of the four GIS models requires information regarding the current extent of the flooding as an input. 
This \textit{flood layer} can originate from remote sensing (e.g., UAV-based \cite{Langhammer2018} or satellite-based \cite{Sivanpillai2020,Wegscheider2013, Roth2025}).
We consider a polygon layer of the flood extent that does not take into account the flood depth. 
This accounts for the fact that inferring the flood depth can be a time-consuming and resource-intensive procedure \citep{Akinboyewa2024}, which makes it impractical for (near) real-time applications. 

In the following, the four GIS models that inform the four leaf nodes (see Figure \ref{fig:bn}) of the area-specific BNs are presented: 

For the \textbf{Density of Exposed Buildings}, first, for each tile, the number of buildings exposed to the flood is assessed using an overlay analysis of the flood extent and the location of buildings (extracted from Open Street Map (OSM) data \citep{OpenStreetMap}). 
Second, the 90\textsuperscript{th} and 75\textsuperscript{th} percentiles of the highest exposed building density are identified, corresponding to the node states \textit{High} and \textit{Medium}, respectively.
Using percentiles to categorize safety-related data is particularly useful when the data does not exhibit clear clusters in its histogram (which, however, can not be expected in this context; see e.g. Fig. \ref{pattern} from the case study).
All remaining exposed tiles correspond to state \textit{Low} and all unexposed tiles correspond to state \textit{None}.
The results are then used as hard evidence to inform the node \textit{Density of Exposed Buildings} of the tile-specific BNs, where the state \textit{None} is assigned to all tiles without any exposed buildings.

The \textbf{Presence of Exposed Care Facilities} is identified by performing an overlay analysis of the flood extent and all care facilities in a tile. 
If an exposed care facility is present in a tile, the corresponding leaf node of the tile-specific BN is set to \textit{True} -- if not, to \textit{False} respectively.

The \textbf{Accessibility of Immediate Unexposed Areas} is estimated by calculating the percentage of the non-flooded and thus unexposed area within the tile and its neighbouring tiles (see Fig. \ref{fig:single_hex}) -- processed as soft evidence in the corresponding tile-specific leaf node.

The \textbf{Accessibility of Remote Unexposed Areas} is modelled using an adapted version of the accessibility model presented in \cite{Schneider2025a}.
To this end, the road network of the study area is reconstructed using OSM data \citep{OpenStreetMap}, resulting in a network topology that features nodes representing road crossings and edges representing road segments \citep{Li2023}. 
Road segments that show an overlay with the flood extent layer are assumed to be flooded and are excluded from the topology. 
In this approach, we neglect the possibility to pass through road segments that are partially flooded \citep{Li2023,Gangwal2022} since we do not consider the flood depth but only the flood extent.
We assume that each remaining road segment can be used for evacuation purposes regardless of any traffic regulations such as one-way streets. 
For the routing algorithm, we defined multiple destination locations (places to escape to) that are sufficiently distant from flooded areas and which are very well connected to the road network to assess whether one can, in general, escape from a tile to a distant, unexposed area via the road network.
To minimise potential biases introduced by the destination location, multiple locations are necessary.
A tile is considered as accessible if a route from at least one unexposed node in the tile to a destination location node exists, i.e. the routing algorithm is applied to each unexposed node in the tile and destination location node.
As a result, each tile is assessed as either \textit{Accessible} or \textit{Inaccessible} informing the node \textit{Road accessibility} of the tile-specific BN (processed as hard evidence).

\subsection{Prioritisation} \label{recommend_method}
To prioritise the areas in need, the probability distributions of the tile-specific target nodes \textit{Risk of People in Need of Assistance} are translated into specific recommendations. 
Recommendations should be based on the entire probability distribution of the target node, including all states \textit{High}, \textit{Medium}, \textit{Low}, and \textit{None}, instead of being only based on the most critical state --  for instance, a probability distribution indicating $0\%$ high risk but $100\%$ medium risk is clearly more concerning than one showing $0\%$ high risk and $100\%$ no risk (see motivation in Section \ref{motivation}).
In a disaster context, even a low level of risk may still necessitate immediate action (i.e. \textit{None} $\ll$ \textit{Low}) and should thus have an influence on the recommendation provided by the method.

In order to consider all target node states, we introduce the \textit{Probability Distribution Criticality} (\textit{PDC}) value 
\begin{equation}
    \textnormal{PDC} = \sum_{i=1}^{N} w_{i}{P(s_{i})} \; ,
\end{equation}
with \textit{PDC} $\in [0, 1]$. The \textit{PDC} combines all target node state probabilities $P(s_{i}) \in [0, 1]$, each with a specific weighting factor that ensures equal distance (in terms of criticality) between the states, i.e. $[w_{None}=0,w_{Low}=0.33,w_{Medium}=0.66,w_{High}=1]$ -- these weighting factors can easily be adjusted, if required, to reflect different preferences of end users.
The closer the \textit{PDC} value is to one, the higher the criticality of the corresponding tile. 

Based on the \textit{PDC} values, all tiles within the study area are assigned to categories of similar risk level. 
These categories constitute the final recommendations for area prioritisation. 
For demonstration purposes, we define four prioritisation categories:
\begin{itemize}
\item[(1)]{\textit{High Priority} tiles constitute the highest risk level of people in need of assistance.}
\item[(2)]{\textit{Priority} tiles also constitute a high, but not the highest, risk level.}
\item[(3)]{\textit{Exposed} tiles contain exposed buildings but do not require immediate prioritisation.}
\item[(4)]{\textit{Safe} tiles are tiles where no exposed people are expected.}
\end{itemize}
The first three of these clusters are assigned based on similar \textit{PDC} values using a $k$-means clustering algorithm, while all tiles that show a \textit{PDC}$=0$ are categorized as \textit{Safe}.

\section{Case Study} \label{case_study}
\begin{figure}[ht]
    \centering
    \includegraphics[width=0.9\textwidth]{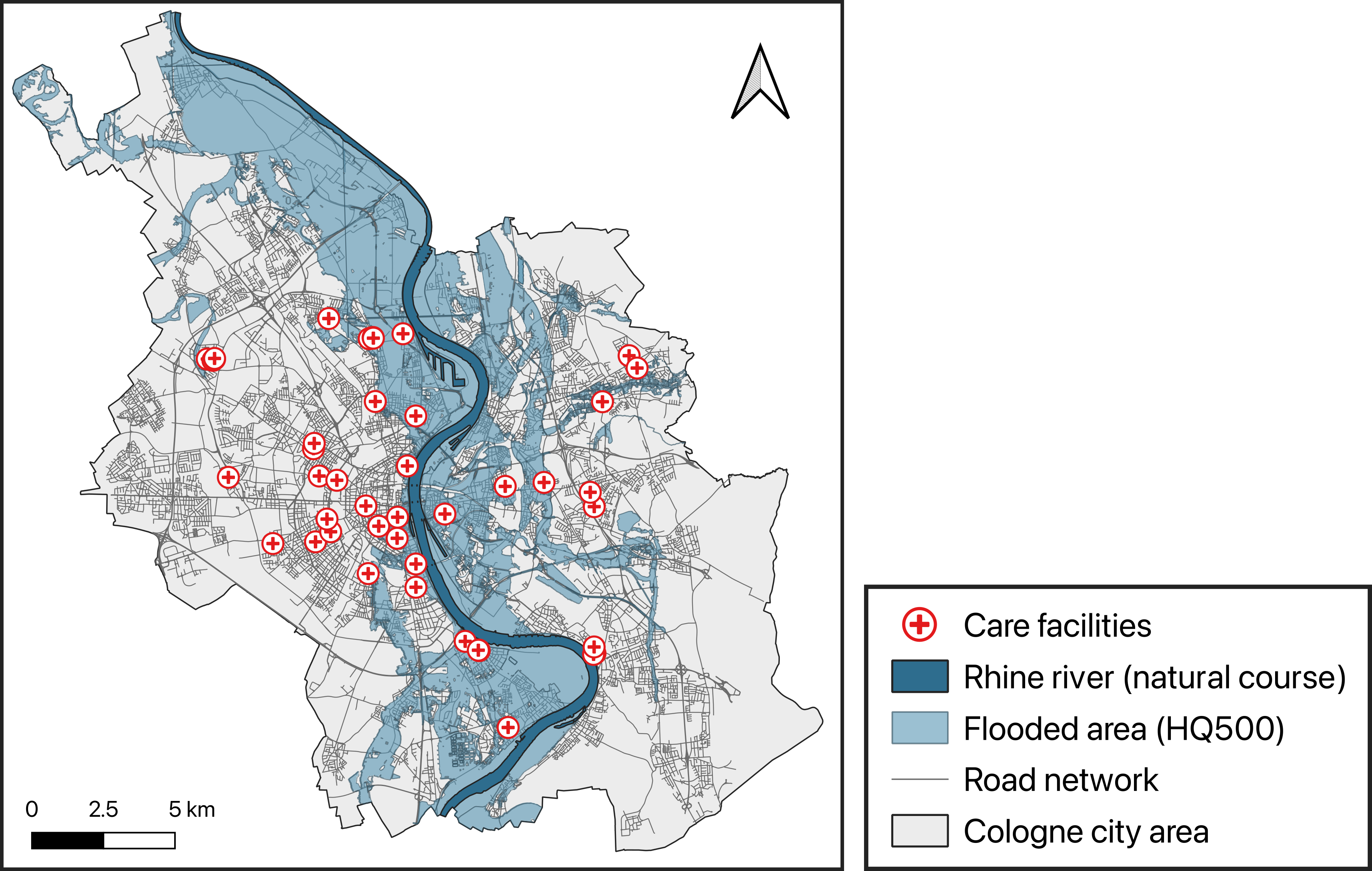}
    \caption{Case study area.}
    \label{fig:case_study_overview}
\end{figure}

We illustrate the \textit{PrioReMap} method in a flood scenario in Cologne, a German city with more than one million inhabitants \citep{Schneider2025a}.
In the past, the city of Cologne has proven to be vulnerable to river floods. 
For instance, in 1993 and 1995, the city experienced floods with severe consequences for its inhabitants and the local economy \cite{Fink1996, nhess-10-1697-2010}.
The city's vulnerability stems from its immediate proximity to the Rhine river which runs right through the densely populated city centre (see Figure \ref{fig:case_study_overview}). Recently, the importance of an effective flood risk management in this region has been demonstrated by the flood disaster in the Ahr Valley in 2021 \cite{Bier2023, Mueller2025}.  

The case study is intended to showcase how the \textit{PrioReMap} can be applied during a flood response. We choose data that resembles the information that is typically available during an ongoing flood. For the flood extent layer, we use data from a hydrological simulation of an extreme flood scenario (also called 500-year flood or HQ500 \citep{HQ500}), which is typically used to create flood risk maps and inform flood protection planning (e.g. see \cite{Fekete2020}).
As this dataset has the same structure as flood extent layers generated by rapid mapping technologies, it could easily be replaced with the most recent snapshot of the flood extent to capture the temporal evolution of flood extent during an actual event.
Furthermore, we use GIS data reconstructed from Open Street Map (OSM) that comprises building locations, care facility locations, and the road network.
As it is reasonable to assume that this information does not change over the course of a flood event, there is no need to replace it with real-time data.
However, the data should be checked for accuracy and completeness before being transferred to a real-world application. 

\subsection{Model results}
\subsubsection{GIS models} \label{spatial_results}
\begin{figure}[ht]
    \centering
    \includegraphics[width=.7\textwidth]{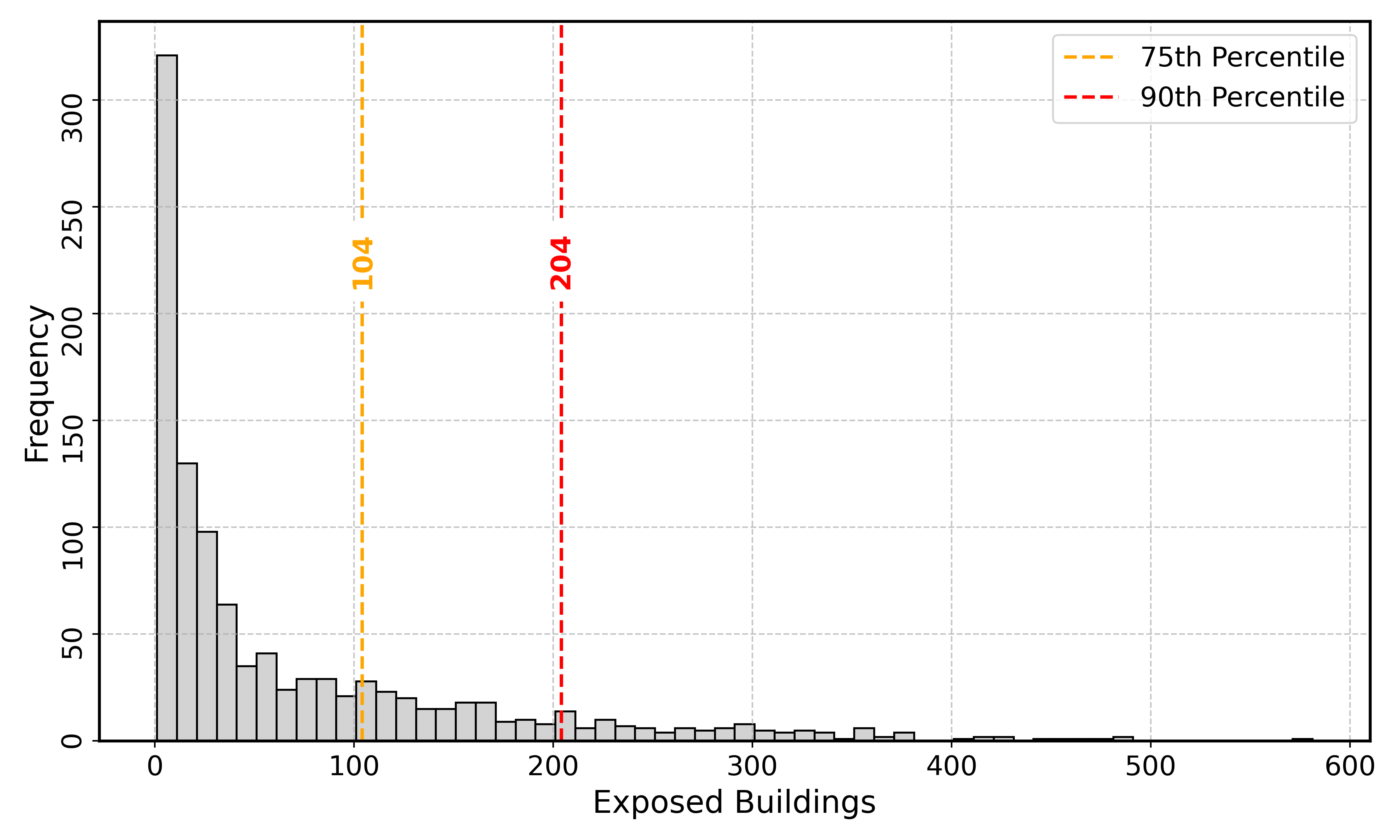}
    \caption{Histogram of exposed buildings per tile.}
    \label{fig:exposed_density}
\end{figure}

\begin{figure}[]
    \centering
    \begin{subfigure}[t]{0.4\textwidth}
        \centering
        \includegraphics[width=\linewidth]{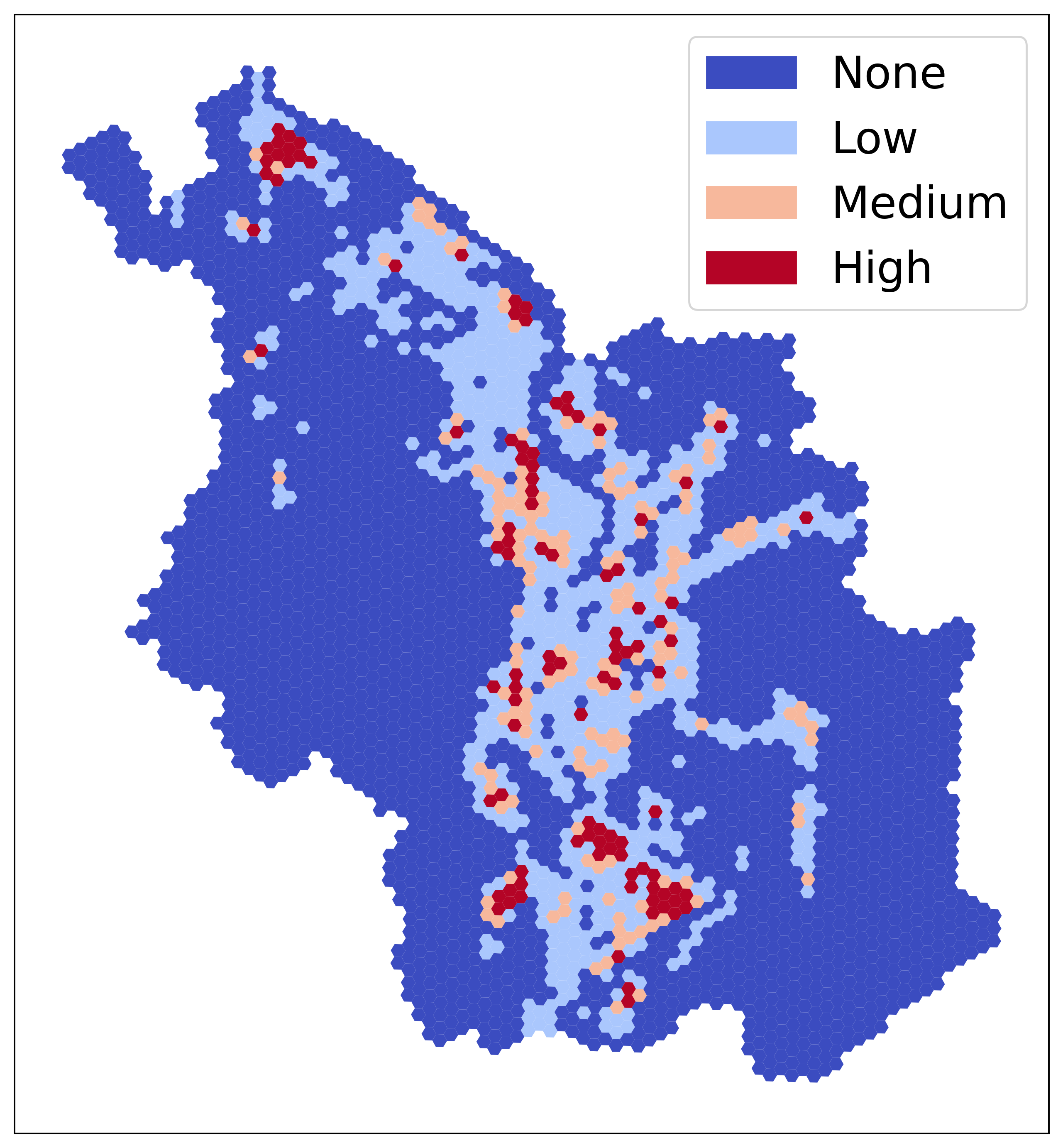}
        \caption{Node \textit{Density of Exposed Buildings}.} \label{building_density}
    \end{subfigure}
    \hfill
    \begin{subfigure}[t]{0.4\textwidth}
        \centering
        \includegraphics[width=\linewidth]{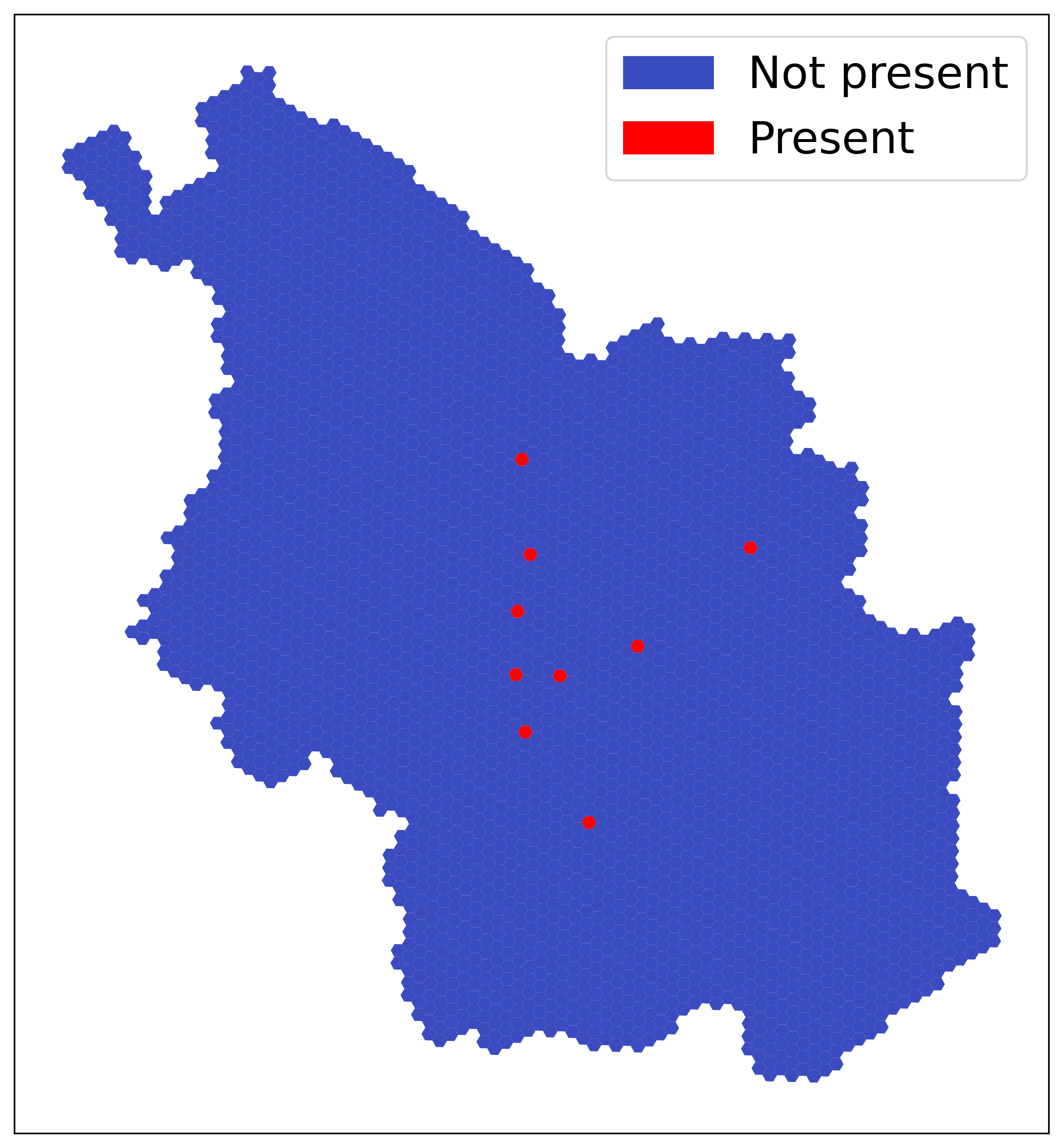}
        \caption{Node \textit{Exposed Care Facilities}.} \label{care_density}
    \end{subfigure}

    \vspace{0.5cm} 

    \begin{subfigure}[t]{0.4\textwidth}
        \centering
        \includegraphics[width=\linewidth]{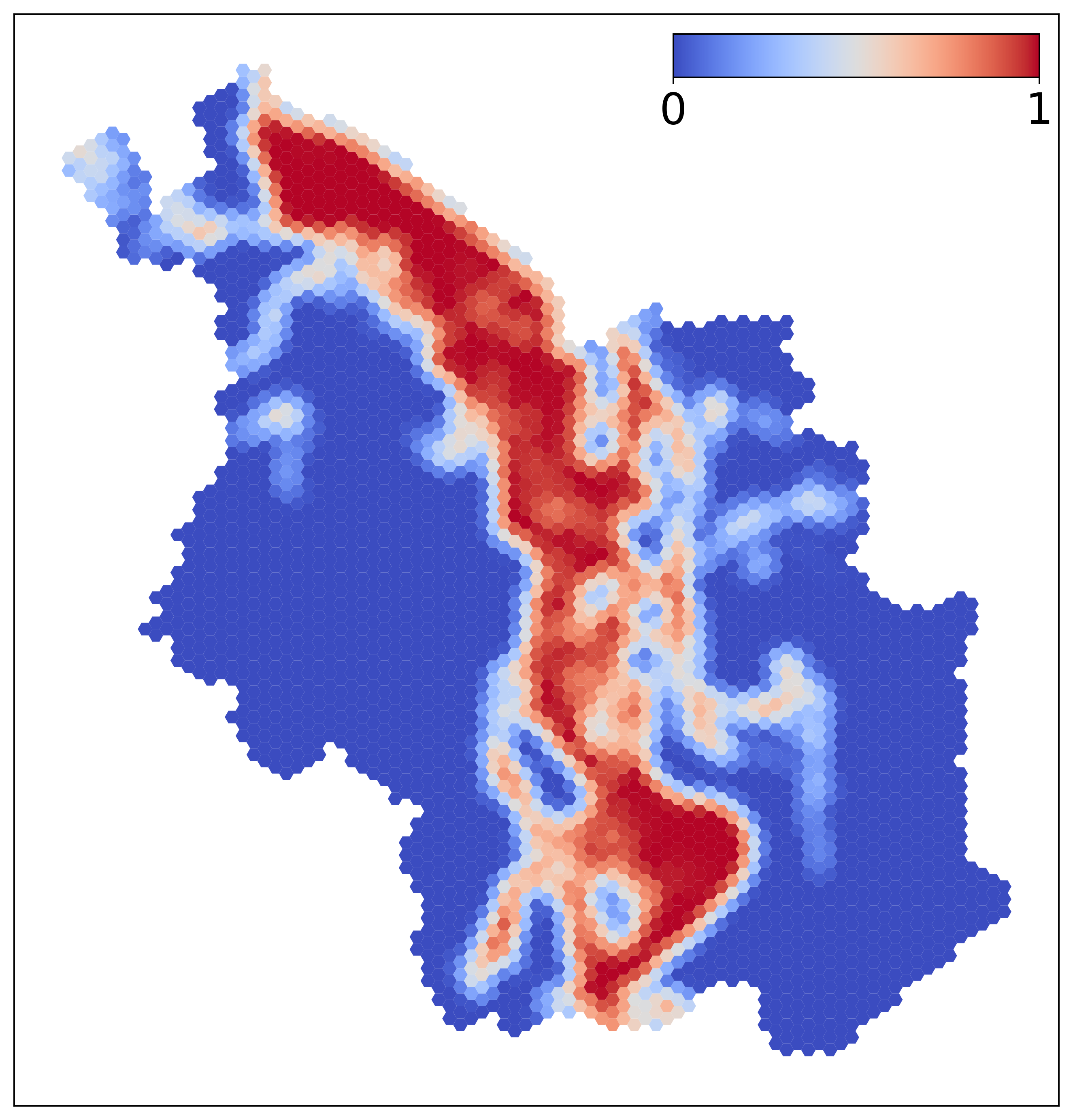}
        \caption{Node \textit{Accessibility of Immediate Unexposed Areas}.} \label{flood_density}
    \end{subfigure}
    \hfill
    \begin{subfigure}[t]{0.4\textwidth}
        \centering
        \includegraphics[width=\linewidth]{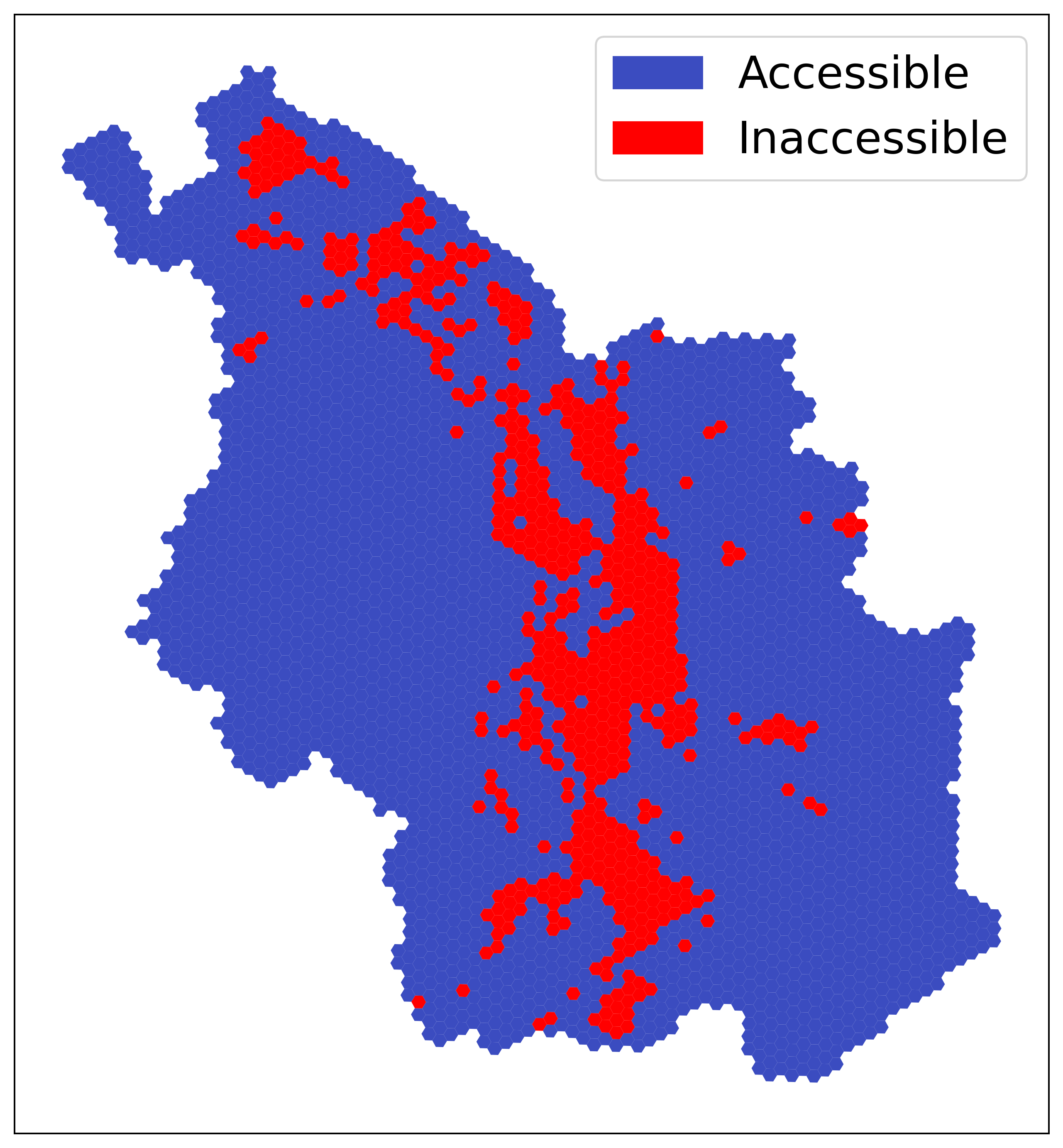}
        \caption{Node \textit{Accessibility of Remote Unexposed Areas}.} \label{access_density}
    \end{subfigure}
    
    \caption{Results of the GIS models.}
\end{figure}

The tile-resolution selected in the case study results in a coverage of $0.114km^2$ ($420m$ in max. width) per tile leading to a total of $3740$ tiles in the study area.
According to the GIS model assessing the \textbf{Density of Exposed Buildings}, $1071$ tiles ($\sim29\%$) contain at least one flooded building.
The categorization based on the 90\textsuperscript{th} and 75\textsuperscript{th} percentiles of the exposed building density distribution (Figure \ref{fig:exposed_density}) rates tiles with $1$ to $104$ flooded buildings as \textit{Low} density, tiles with $104$ to $203$ as \textit{Medium} density, and tiles with more than $203$ flooded buildings as \textit{High} density (see Figure \ref{building_density}). 
In a total of $9$ out of $39$ tiles that contain care facilities, at least one \textbf{Exposed Care Facility} is present (Figure \ref{care_density}).
To assess the \textbf{Accessibility of Immediate Unexposed Areas}, the flood density within each tile and its neighbouring tiles is calculated.
This leads to the identification of $1713$ tiles in which the immediate neighbourhood is only partly accessible, with flood coverage varying between $\sim1\%$ and $\sim99\%$ (see Figure \ref{flood_density}).
To determine the \textbf{Accessibility of Remote Unexposed Areas}, the routing algorithm is applied, resulting in the identification of $614$ tiles that are rendered inaccessible by the road network due to flooded road segments (see Figure \ref{access_density}).

\subsubsection{GIS-informed BN models} \label{spatial_bn_results}
\begin{figure}[]
    \centering
    \begin{subfigure}[t]{0.45\textwidth}
        \centering
        \includegraphics[width=\linewidth]{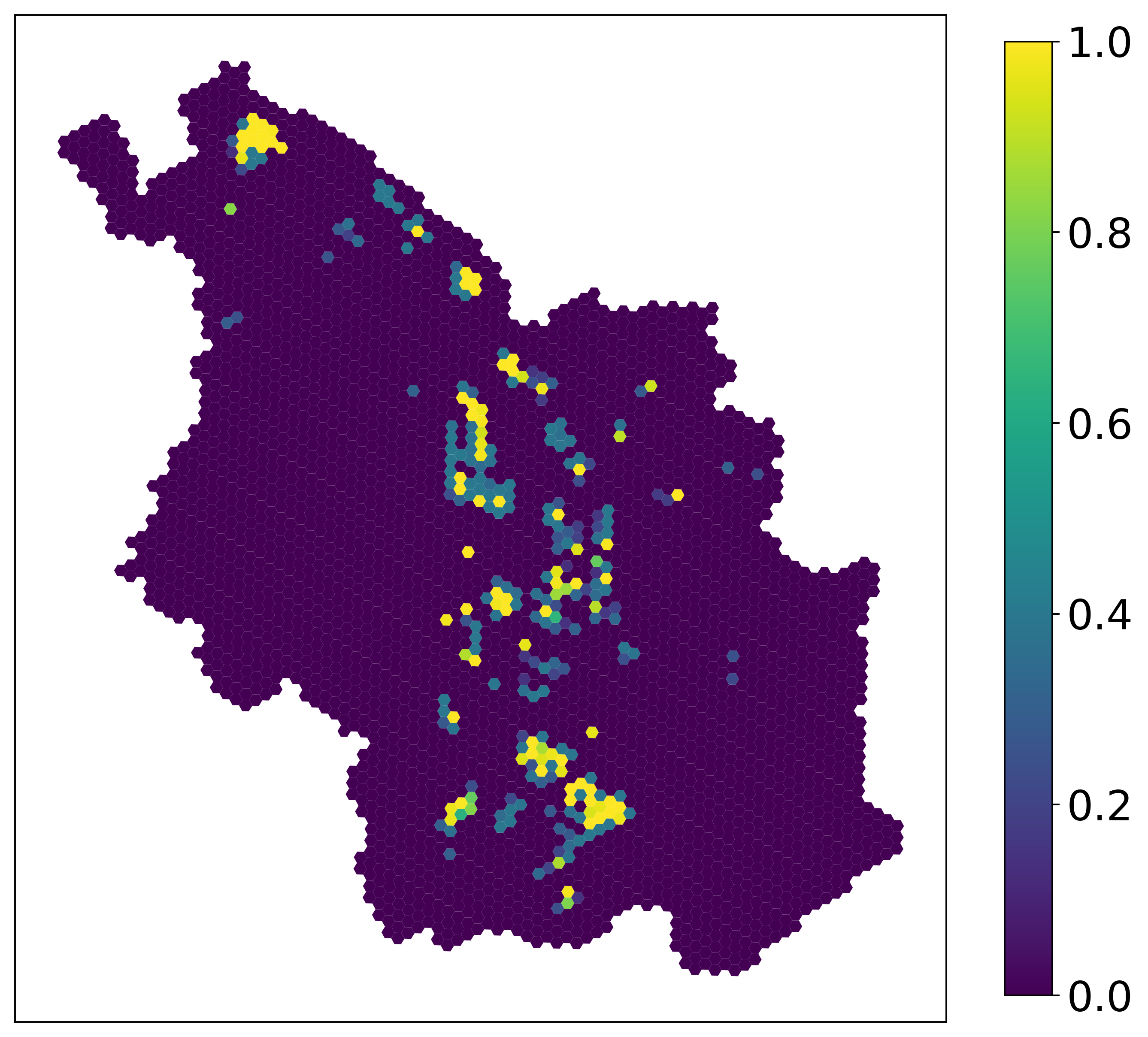}
        \caption{State \textit{High}} \label{high_map}
    \end{subfigure}
    \hfill
    \begin{subfigure}[t]{0.45\textwidth}
        \centering
        \includegraphics[width=\linewidth]{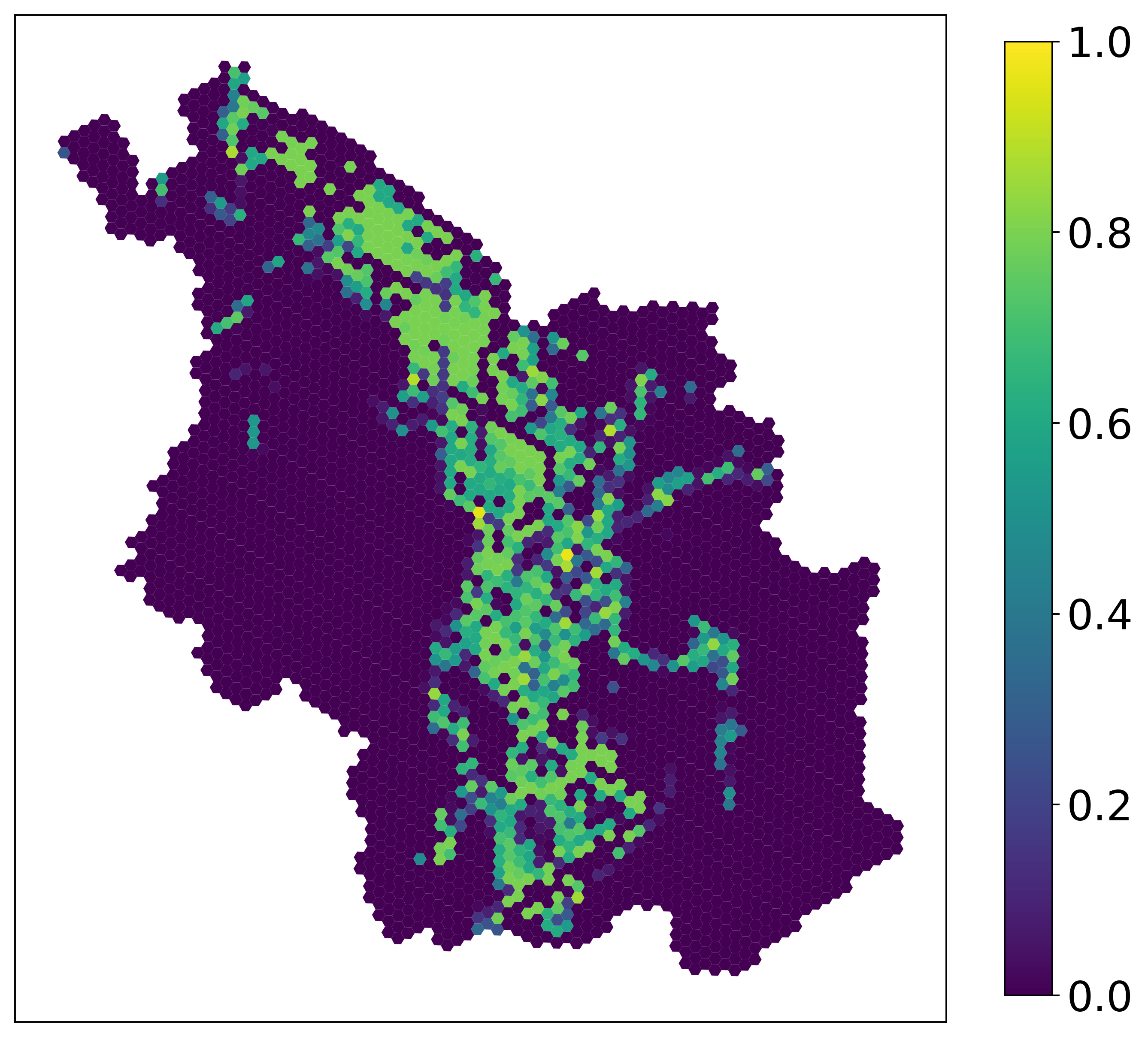}
        \caption{State \textit{Medium}} \label{medium_map}
    \end{subfigure}

    \vspace{0.5cm} 

    \begin{subfigure}[t]{0.45\textwidth}
        \centering
        \includegraphics[width=\linewidth]{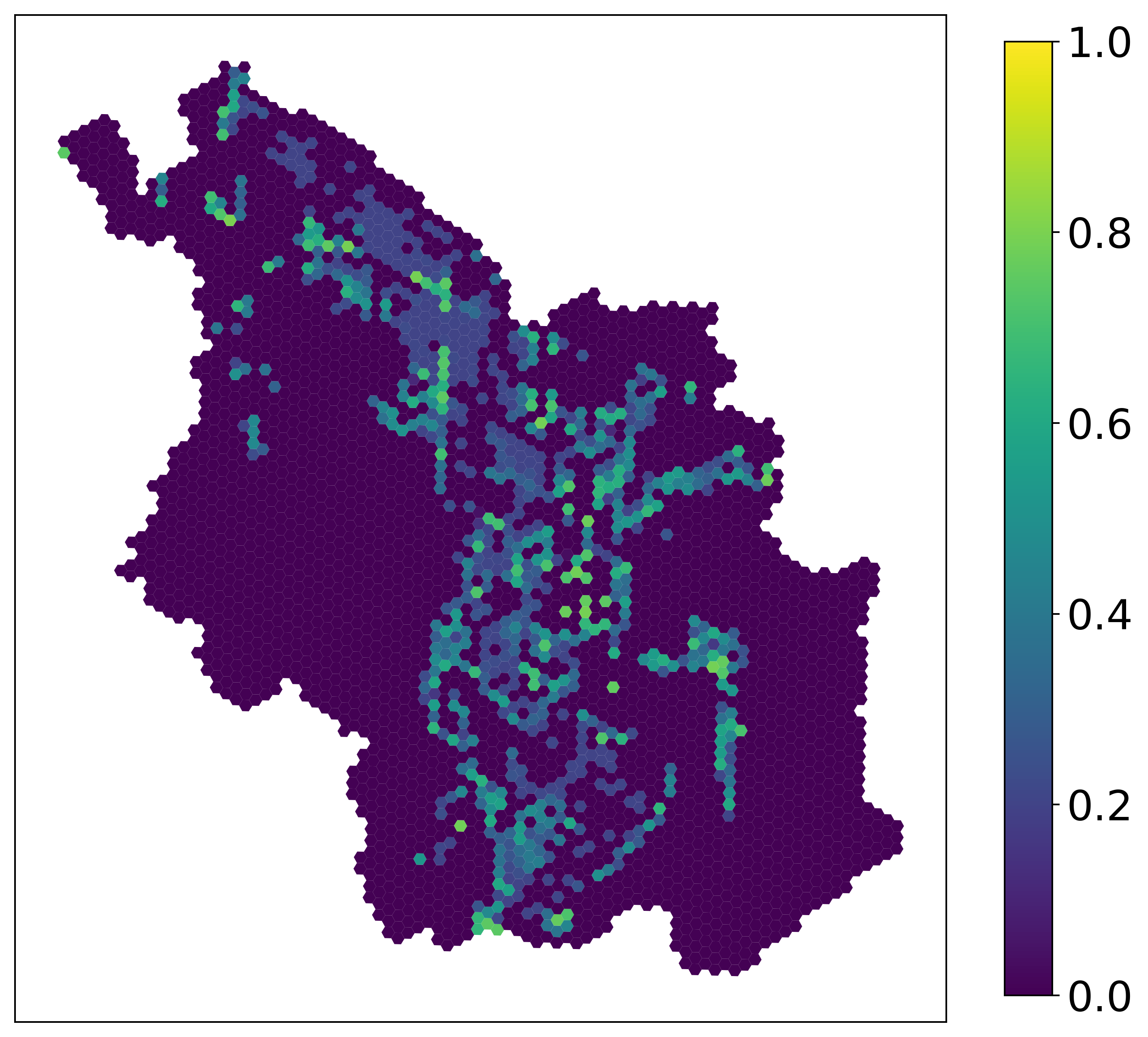}
        \caption{State \textit{Low}} \label{low_map}
    \end{subfigure}
    \hfill
    \begin{subfigure}[t]{0.45\textwidth}
        \centering
        \includegraphics[width=\linewidth]{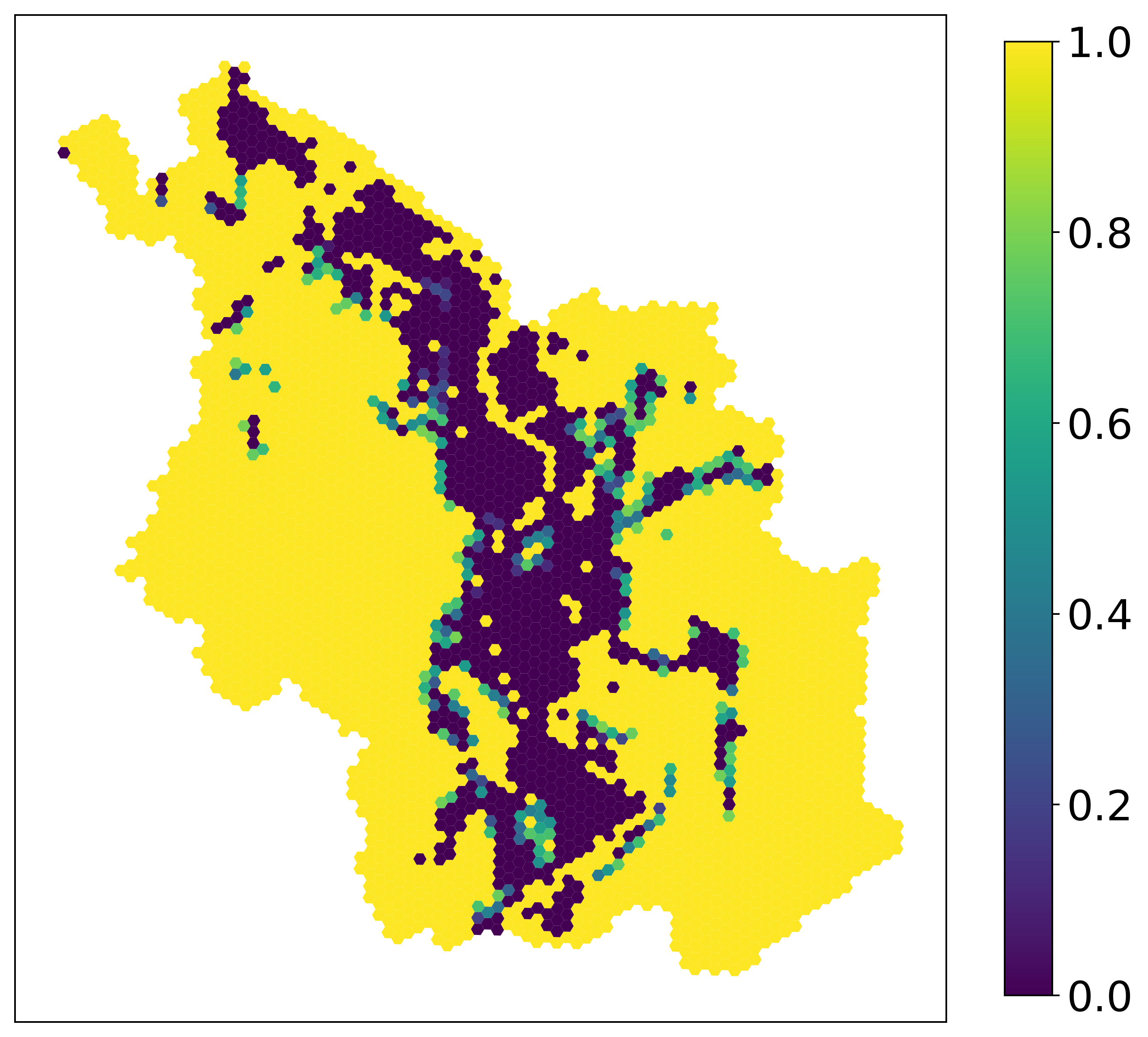}
        \caption{State \textit{None}} \label{none_map}
    \end{subfigure}
    
    \caption{Results of the GIS-informed BN model.}
\end{figure}

Using the results of the GIS models (see Section \ref{spatial_results}) as inputs for the tile-specific BNs (see Section \ref{bn_model}), the probability distribution for the states of the target node \textit{Risk of People in Need of Assistance} is calculated for each tile in the study area. 
From the total of $3740$ tiles, $1071$ tiles ($\sim30\%$) are exposed, i.e. they are (partially) flooded and contain at least one exposed building. 
A total of $32$ tiles ($\sim1\%$) show the highest risk, i.e. state \textit{High} is equal to one (see Figure \ref{high_map}), nine of which include an exposed care facility which sets their probability of being in the \textit{High} risk state directly to one.
The remaining $23$ maximum risk tiles are completely flooded, exhibit a high density of exposed buildings, and do not allow the accessibility of immediate or remote unexposed areas.
The remaining $1039$ exposed tiles show a probability distribution without extreme values, resulting from a mix of critical and less critical node states, e.g., tiles with a medium density of flooded buildings that remain accessible, or tiles with a low density of flooded buildings that are rendered inaccessible.

\subsection{Area prioritisation}
\begin{figure}[ht]
    \centering
    \includegraphics[width=0.8\textwidth]{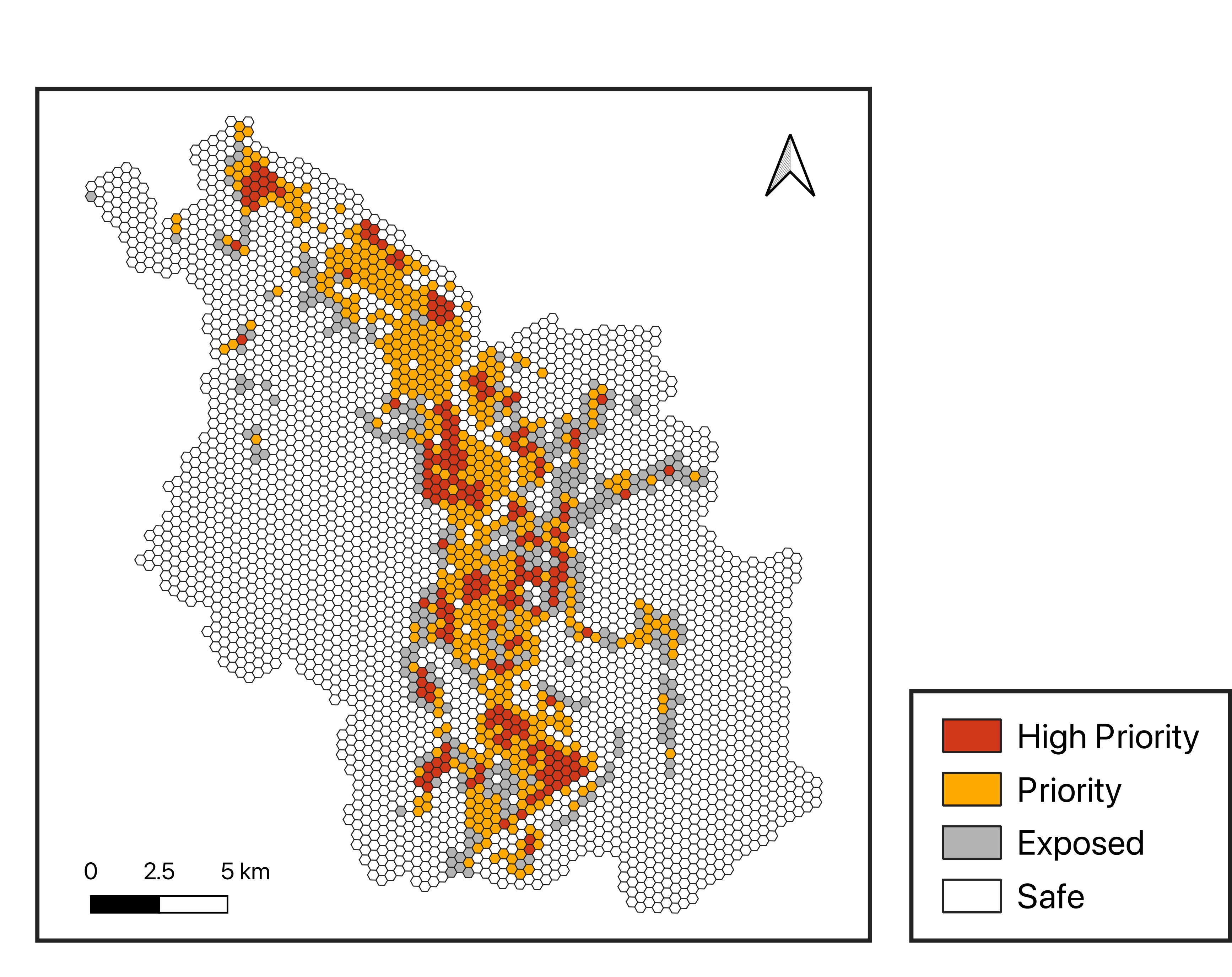}
    \caption{\textit{PrioReMap} of the case study.}
    \label{prioremap}
\end{figure}

Applying the recommendation method (see Section \ref{recommend_method}) to the results of the GIS-informed BNs (see Section \ref{spatial_bn_results}) concludes in the final result of the method: the \textit{PrioReMap} (Figure \ref{prioremap}). This map displays the recommendations for area prioritisation in the study area according to the prioritisation classes \textit{high priority}, \textit{priority}, \textit{exposed}, and \textit{safe} (see Section \ref{recommend_method}). 

\begin{figure}[ht]
    \centering
    \includegraphics[width=0.8\textwidth]{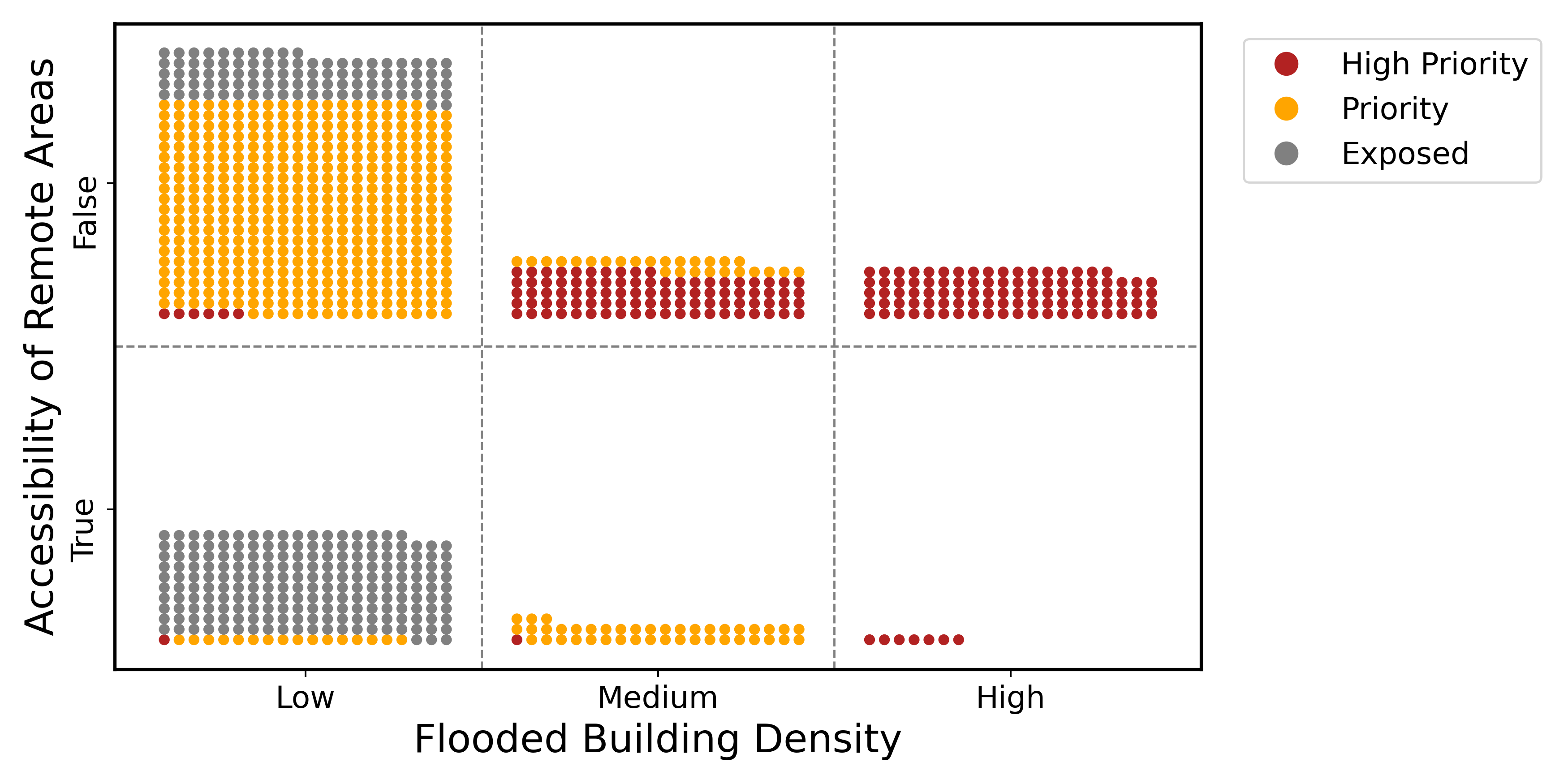}
    \caption{Tile similarities based on leaf node states.}
    \label{pattern}
\end{figure}

Overall, $202$ tiles ($\sim5\%$ of the study area) are categorized as \textit{high priority} areas.
The \textit{PrioReMap} reveals multiple spatial clusters of \textit{high priority} tiles (e.g. the cluster at the top of Fig. \ref{prioremap} consisting of $14$ tiles), as well as several isolated \textit{high priority tiles} -- often, but not always, associated with the presence of exposed care facilities.
These clusters are not necessarily situated close to the natural course of the Rhine river. For example, the \textit{high priority} cluster in the bottom left of the \textit{PrioReMap}, which consists of nine tiles, is located approximately $2.5km$ from the river’s natural course.
In general, there are three constellations of BN leaf node states in the category of \textit{high priority} tiles (that result from the clustering of the \textit{PDC} values, as introduced in Section \ref{recommend_method}):
(i) an exposed care facility is present in a tile (regardless of flooded building density and accessibility);
(ii) a tile shows a high density of flooded buildings (right row in Figure \ref{pattern}); and
(iii) a tile shows a medium density of flooded building, is inaccessible from remote areas and has a high percentage ($>75\%$) of flooded areas in its immediate vicinity (second row, first line in Figure \ref{pattern}).

A total of $536$ tiles ($\sim14\%$ of the study area) are categorized as \textit{priority} areas.
\textit{Priority} tiles are also distributed throughout the exposed study area. They often surround clusters of \textit{high priority} tiles (e.g. at the top of Fig. \ref{prioremap}) but also occur as isolated instances.
These areas exhibit three constellations of BN leaf node states:
a tile shows a medium density of exposed buildings and is either (i) accessible by remote areas or (ii) inaccessible by remote areas and has less than $75\%$ of flooded areas in its immediate vicinity (middle row in Figure \ref{pattern});
a tile shows a low density of exposed buildings and is either (iii) accessible by remote areas and shows more than $80\%$ of immediate flooded areas or (iv) inaccessible by remote areas and shows more than $50\%$ of immediate flooded areas (left row in Figure \ref{pattern}).

A total of $333$ tiles ($\sim9\%$ of the study area) are categorized as \textit{exposed} areas.
Tiles categorized as \textit{exposed} -- those that contain exposed buildings but do not require immediate prioritisation -- often appear along the edges of the flood inundation, as these areas are more likely to be accessible from both immediate and remote non-flooded regions.
These areas exhibit two constellations both of which include a low density of exposed buildings:
(i) accessible by remote areas and shows less than $80\%$ of immediate flooded areas or
(ii) inaccessible by remote areas and shows less than $50\%$ of immediate flooded areas (left row in Figure \ref{pattern}).

\section{Discussion and Conclusion} \label{discussion}
In this paper, we introduce \textit{PrioReMap}, a novel method to provide area prioritisation recommendations in flood disaster response. 
Our method consists of three core elements.
First, a grid of BNs is used to infer the spatially distributed \textit{Risk of People in Need of Assistance} based on spatial variables, such as the presence of exposed care facilities and the accessibility of unexposed areas. 
Second, GIS models that inform the tile-specific BN leaf nodes using a flood inundation layer as hazard map.
Third, a prioritisation method that translates the tile-specific probability distributions of the \textit{Risk of People in Need of Assistance} into distinct prioritisation categories.

A central motivation of our \textit{PrioReMap} method is enabling rapid and transparent area prioritisation to support disaster decision making in an ongoing flood. 
%
To achieve rapid recommendations, the method shifts preparation for the prioritisation from the time-critical response phase to the less-pressured preparedness phase. 
During the preparedness phase, setting up the models (GIS models and BN) effectively serves as a walk-through of the decision-making process -- identifying relevant variables for prioritisation and establishing the logical dependencies between these variables and the objective (by building the BN model), i.e. the \textit{Risk of People in Need of Assistance}.
This approach is particularly valuable when integrating knowledge from multiple experts into the models.
Often, selection problems in emergency decision making are solved by group decisions \citep{Zheng2020} and are mostly done manually through document editors, or even verbally \citep{SkoeldGustafsson2023}.
In addition, setting up the models promotes a clear understanding of how the method generates its recommendations -- especially in high stake decisions, there is a reluctance to trust black box models \citep{Waal2022} and ease of interpretation is vital \citep{lee2022roadmap}.
Bayesian networks can be readily derived from expert knowledge (e.g. see \cite{Hassall2019} or \cite{Morris2014} for elicitation methods) and effectively reflect the cognitive processes of decision makers 
(\cite{peal1985bayesian} introduced BNs as a formalism of human reasoning under uncertainty)
-- precisely aligning with our goal of formalising decision making during the preparedness phase to facilitate rapid and transparent results during the response phase.

However, deriving a single value (the area recommendation) from multiple variables (the BN-leaf nodes) is a sensitive task \citep{Halekotte2025}.
%
We selected a BN as the core of our method for aggregating geospatial variables, as the effectiveness of this approach has already been demonstrated in similar contexts, e.g. in flood disaster preparedness (see Table \ref{studies}). 
A key advantage of BNs is their ability to model the complex dependencies between variables through CPTs, which capture all combinations of dependent node states. 
To reduce the effort required for knowledge elicitation during BN construction and to significantly limit the number of necessary probability values, we discretized the variable \textit{Density of Flooded Buildings} into four states (\textit{None}, \textit{Low}, \textit{Medium}, \textit{High}) based on percentiles. 
While this approach introduces threshold-based classifications that certainly influence the results, it also provides a mechanism to adjust the number of high priority and priority areas -- for instance, shifting the percentile thresholds towards higher percentiles reduces the number of areas classified as high priority.

In the presented use case, the geospatial data that is required to set up the GIS models is obtained from OSM data. 
However, in this context, data from official government services or critical infrastructure providers should generally be preferred, as these datasets are typically more accurate and complete \citep{Schneider2025a}.
When relying on OSM data, potential inaccuracies -- such as missing, misplaced, or incorrectly labelled geo-information -- must be taken into account \cite{Kaur2017}. 
To address these issues, existing methods for classifying and improving the quality of OSM data can be applied (see, e.g. \cite{Brovelli2018} or \cite{Biljecki2023}).
Ultimately, it is important to emphasise that the effectiveness of \textit{PrioReMap} is directly tied to the quality of the underlying data, as accurately assessing accessibility in remote areas depends on a reliable reconstruction of the road network.
However, potential end users, i.e. disaster responders, may already possess crucial data of higher quality than what is publicly available.

For the flood inundation layer, we demonstrated our method in a case study using a layer generated from a simulated flood scenario (HQ500). This layer has the same structure as a layer produced by rapid mapping technologies, i.e. a polygon of the flood extent.
While flood inundation is inherently a dynamic phenomenon, our approach simplifies it as a static snapshot for the presented case study (e.g. see \cite{unspider_flood_mapping_s2} for a step-by-step guide on generating such layers using open-source data and tools), but the approach can be easily adjusted via updating floodmaps at run-time as new information is provided. 
In addition, incorporating flood depth into the analysis could enhance the accuracy of the results, for instance to better assess accessibility. At the same time, we seek to balance accuracy with speed of assessment, and flood depth is not easily accessible (since a digital elevation model of the area is required) and not rapidly available (because of high computational costs). 
In the future, AI-based methods, such as the one proposed by \cite{Akinboyewa2024}, might allow for more rapid and reliable inference of flood depth, making its integration increasingly feasible.

We identified two directions for future research. 
First, additional observations of ongoing flood disasters, such as eyewitness reports, emergency calls, or sensor information, could be included in the analysis. 
To achieve this, the method could be combined with the \textit{Emergency Response Inference Mapping} method (see \cite{Schneider2025}), which allows for the integration of such dynamic and potentially uncertain observations. The integration of such observations should be carefully discussed with potential end users, as relying on them for decision making may introduce new uncertainties and hamper accountability. For instance, the absence of emergency calls from a particular area does not necessarily indicate lower urgency -- people in those areas may simply lack mobile coverage or access to other communication channels.
Second, the method could be extended to determine sectors, i.e. spatial clusters that balance geographic size and the number of (high) priority tiles, to support emergency relief efforts. 
Sectorisation is a common procedure in such contexts (e.g. see \citep{INSARAG2020}). 
In addition, for each sector, appropriate locations, such as large parking lots or other public buildings, can be identified to serve as temporary hubs for coordinating response personnel and distributing materials.


\end{document}